\documentclass[3p, 11pt, onecolumn]{elsarticle}
\usepackage{graphicx,amscd,amsmath,amssymb,verbatim}
\usepackage{amsfonts,epsfig}
\usepackage{mathptmx}
\usepackage{setspace}
\usepackage{epsfig}
\usepackage{url}
\usepackage{multirow}
\usepackage{subfigure}
\usepackage{graphicx,color}
\usepackage{algorithm,caption}
\usepackage{algpseudocode}
\usepackage{float}
\usepackage[normalem]{ulem}

\pdfoutput=1

\begin{document}

\journal{Journal of Statistical Mechanics}

\begin{frontmatter}

\title{Estimating time constants of the RTS noise in semiconductor devices: a complete description of the observation window in the time domain}

\author{Roberto da Silva, Gilson Wirth}

\address{Institute of Physics, Universidade Federal do Rio Grande do Sul,\\
Av. Bento Gon{\c{c}}alves, 9500 - CEP 91501-970, Porto Alegre, Rio Grande do Sul, Brazil\\
}

\address{Electrical Engineering  Department, Universidade Federal do Rio Grande do Sul Porto Alegre, Brazil, Porto Alegre, Rio Grande do Sul, Brazil\\
}

\begin{abstract}

 We obtained a semi-analytical treatment obtaining estimators for the sample variance and variance of sample variance for the RTS noise. Our method suggests a way to experimentally determine the constants of capture and emission in the case of a dominant trap and universal behaviors for the superposition from many traps. We present detailed closed-form expressions corroborated by MC simulations. We are sure to have an important tool to guide developers in building and analyzing low-frequency noise in semiconductor devices. 
 
\end{abstract}

\end{frontmatter}

\section{Introduction}

Knowing the so-called low-frequency (LF) noise \cite{Kirton,Weissman} means
understanding the stochastic process of the capture/emission mechanisms by
traps. Such phenomena can be due to one only single dominant trap or
dominated by multiple random telegraph signals (RTS) due to many traps, in
semiconductor-dielectric interfaces found in CMOS transistors \cite%
{Machlup,SilvaPhysicaA}. Both situations are important and their study is
essential, and of technological interest. To model such a process, we can
imagine a straightforward mechanism described by Fig. \ref%
{Fig:transistor_and_voltage} (a)

In this scheme, a trap captures one charge carrier between the time $t$ and $%
t+\delta t$, with probability%
\begin{equation}
p(0\rightarrow 1)\delta t\approx \frac{1}{\tau _{c}}\delta t\text{.}
\label{Eq:01}
\end{equation}%
This charge carrier returns to the inversion layer with a probability%
\begin{equation}
p(1\rightarrow 0)\delta t\approx \frac{1}{\tau _{e}}\delta t  \label{Eq:10}
\end{equation}%
at the same time interval. Here $\tau _{c}$ and $\tau _{e}$ are capture and
emission constants, respectively. Generally, we make $\delta t=1$ u.t. (unit
of time), imagined as a minimal quantity in a typical Monte Carlo (MC)
simulation, or simply 1 MC step. When a trap captures a charge carrier, one
observes a fluctuation of magnitude $\Delta v$ in the voltage, and one
subtracts this same value in the emission of this charge carrier to the
inversion layer.

\begin{figure}[tbp]
\begin{center}
\includegraphics[width=0.5\columnwidth]{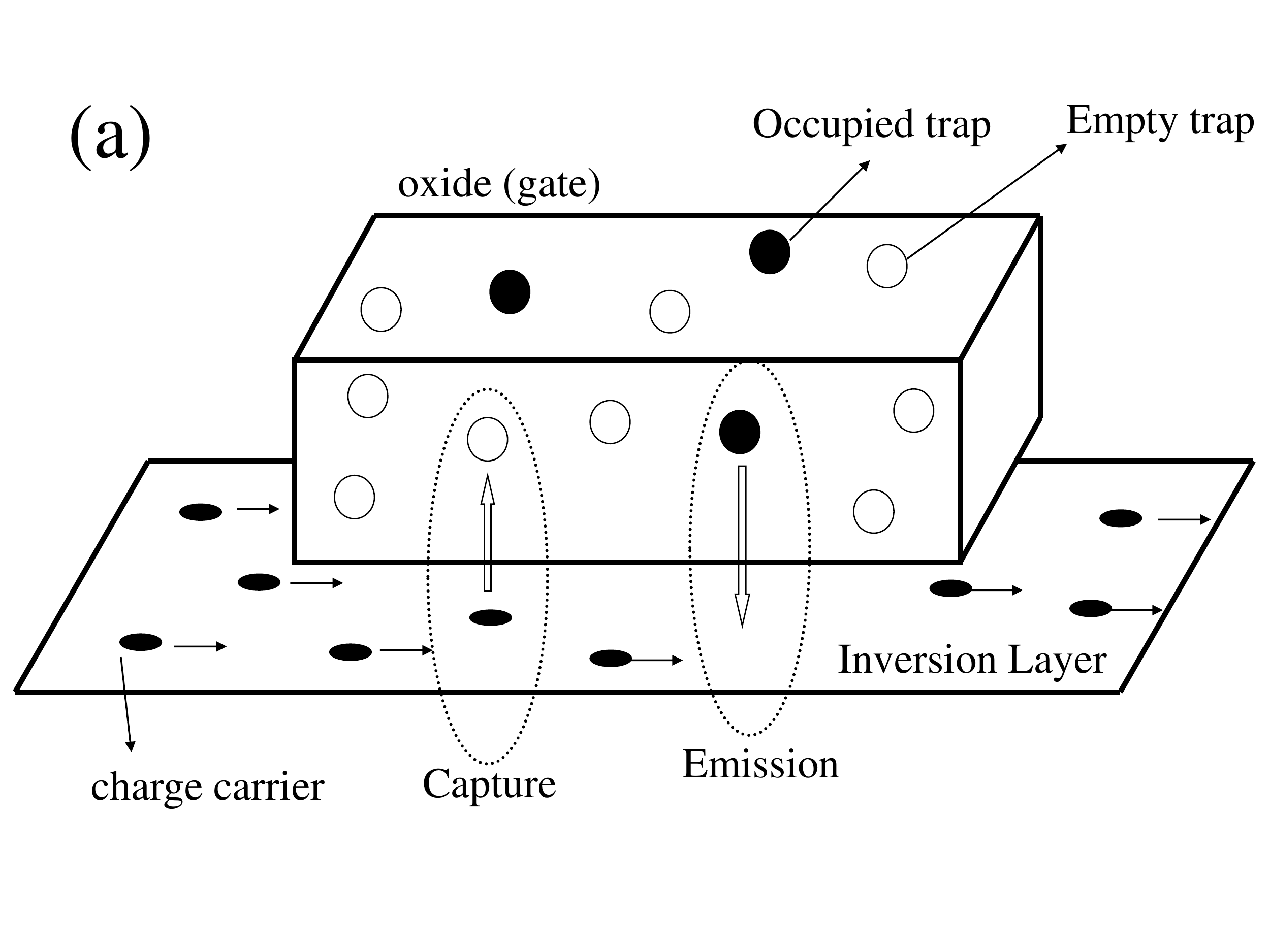}%
\includegraphics[width=0.5\columnwidth]{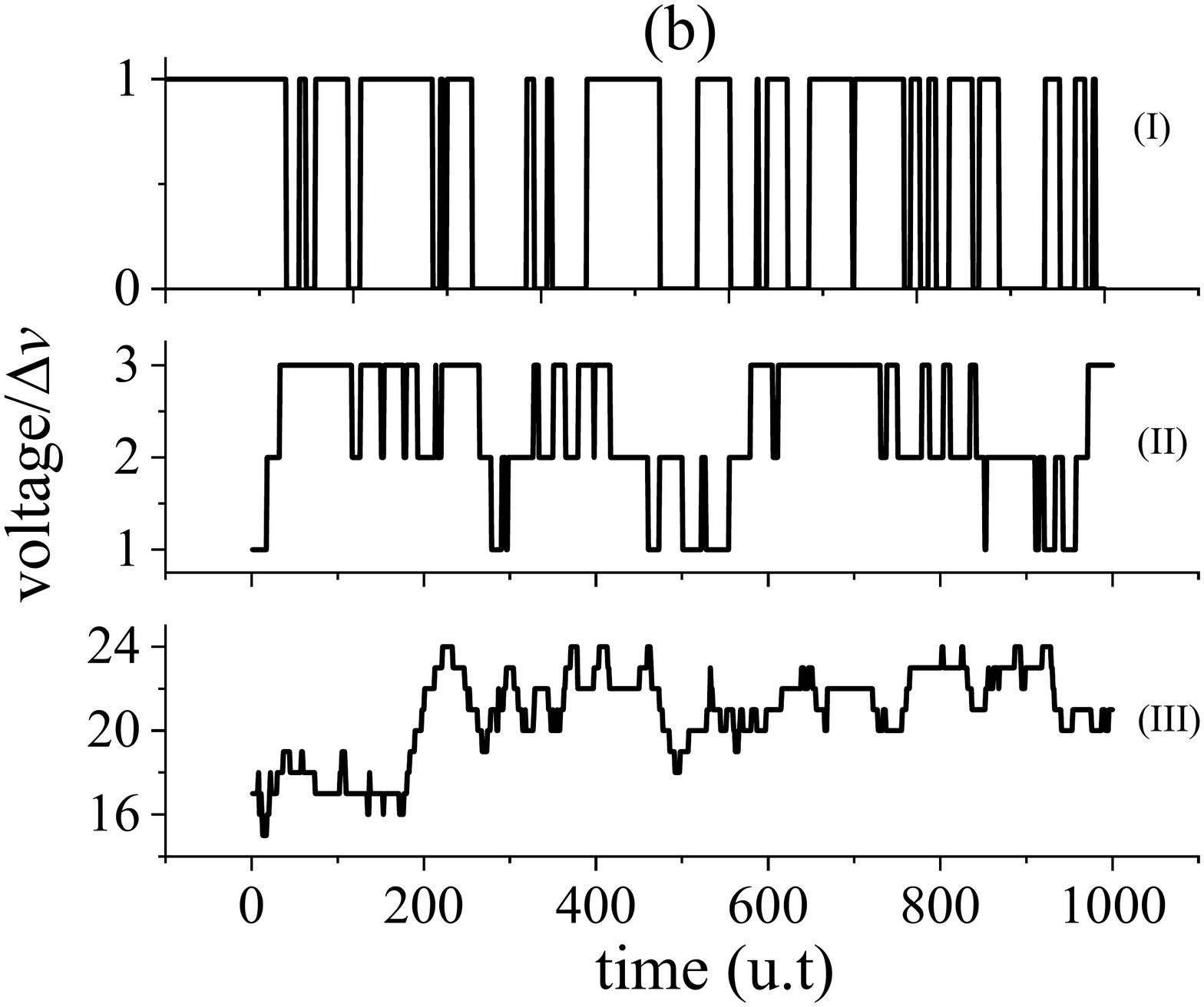}
\end{center}
\caption{(a) A simplification of a CMOS transistor (b) Simulation of the
threshold voltage in a CMOS dispositive considering 1, 3, and 50 traps,
represented by (I), (II), and (III) respectively. }
\label{Fig:transistor_and_voltage}
\end{figure}

Fig. \ref{Fig:transistor_and_voltage} (b) shows the threshold voltage
considering the contribution of one, three, and fifty traps considering
simple MC simulations following the prescription that we previously defined.
In this case, we consider the amplitude $\Delta v$ is the same for all traps
in this simple experiment. We can observe that the sum of these fluctuations
compose exciting patterns, and its understanding is essential to developing
reliable devices.

The phenomenology of random telegraph signals considers that the
distributions: 
\begin{equation}
P_{0(1)}(t)=\frac{1}{\tau _{e(c)}}e^{-t/\tau _{e(c)}}
\end{equation}%
describe the probability density functions which govern the residence time
on the different states $\sigma _{i}=0(1)$, captured/emitted respectively,
where the time constants:%
\begin{equation}
\tau _{e(c)}=\int\limits_{0}^{\infty }tP_{0(1)}(t)dt\text{,}
\end{equation}%
are here translated as averaged capture time ($\tau _{c}$), and averaged
emission time ($\tau _{e}$) respectively.

Although many works consider the analysis in the frequency domain of RTS (to
cite a few ones \cite{Brederlow,Gilson2007,vanderwel,silvajstat,Roy,AMM}),
on the other hand, time-domain analysis (see for example \cite%
{timedomain1,timedomain2,timedomain3,timedomain4}) deserves more attention
from researchers in this kind of modeling. Time-domain analysis is important
for example to understand the degradation phenomena in semiconductor devices
from experimental \cite{Grasser2009}, and theoretical point of view \cite%
{Silva-jstat-2}. However, analysis in the time domain does not explore
studies about the observation window, which seems to be an important
parameter  \cite{Observation,Observation2}.

In this paper, we develop a very detailed approach to study, in time domain
regime, ways to characterize the RTS noise using a semi-analytical treatment
and Monte Carlo (MC) simulations. We first developed an approach for the
case of a single dominant trap. We analyzed the noise variance per time
window and the variance of the sample variance as a function of time window
size. This analysis allows developers, for example, to estimate the
constants $\tau _{c}$ and $\tau _{e}$ independently, since we present
closed-form equations for these two quantities based on the hypothesis that
MC simulations, supposedly imitating an experiment, supply as input the
steady-state of the variance and the point that maximizes the variance of
the variance.

In addition, we also performed an extrapolation of this result for the
contribution of many traps. We show that variance per window shows a
universal linear behavior, as suggested by previous contributions \cite%
{Observation,Observation2}. However these works import previous results
obtained in frequency domain \cite{SilvaPhysicaA} and therefore with
parameters not directly extracted.

We performed an analysis entirely made in the time domain in this current
contribution, which leads to a universal behavior with easily checked
parameters and that presents an excellent agreement with MC simulations.

Finally, we also show a universal behavior for the variance of the variance,
suggesting a regime where this amount does not depend on the time window, a
result that literature never observed.

In the next section, we present our semi-analytical modeling. The
penultimate section presents some results by comparing MC simulations with
these semi-analytical predictions. Finally, in the last section, some
summaries and conclusions are presented.

\section{Semi-analytical predictions}

In this section, we will deduce some formulas for the sample variance and
for the variance of the sample variance, which are used to characterize RTS
and mainly to allow that we can compute the constants $\tau _{c}$ and $\tau
_{e}$. We first prepare an analysis for one only dominant trap. After, we
deduce some expressions considering the contribution of many traps for the
noise.

\subsection{Single trap}

For a single trap, we calculate the threshold voltage average as: 
\begin{equation}
\left\langle v\right\rangle =p_{0}\cdot 0+p_{1}\delta v\text{.}
\end{equation}%
Here $p_{0}$ denotes the probability that an electron is not trapped, while $%
p_{1}$ is the probability that an electron is trapped. In this situation,
there is a contribution of $\delta v$ to the voltage, while we consider that
in the first situation, the voltage variation is 0. Thus $s=\frac{v}{\Delta v%
}$ is a Bernoulli random variable, but only when the observation time of the
noise: $T>>\tau _{c},\tau _{e}$, since, in this scale, $p_{0}$, and $p_{1}$
are given by:%
\begin{eqnarray}
p_{0} &=&\frac{\tau _{e}}{\tau _{e}+\tau _{c}} \\
&&  \notag \\
p_{1} &=&\frac{\tau _{c}}{\tau _{e}+\tau _{c}}\text{,}  \notag
\end{eqnarray}%
such that $p_{0}+p_{1}=1$.

Thus, $\left\langle v\right\rangle =\frac{\tau _{c}}{\tau _{e}+\tau _{c}}%
\delta v$. The dispertion (variance) of the voltage change \ is so
calculated by first calculating the second moment: 
\begin{equation}
\left\langle v^{2}\right\rangle =0p_{0}+\frac{\tau _{c}}{\tau _{e}+\tau _{c}}%
(\delta v)^{2}
\end{equation}%
which yields to $var(v)=$ $\left\langle v^{2}\right\rangle -\left\langle
v\right\rangle ^{2}=\frac{\tau _{e}\tau _{c}}{\left( \tau _{e}+\tau
_{c}\right) ^{2}}(\delta v)^{2}$. We denote:%
\begin{equation}
var_{\infty }=\frac{var(v)}{(\delta v)^{2}}=\frac{\tau _{e}\tau _{c}}{\left(
\tau _{e}+\tau _{c}\right) ^{2}}
\end{equation}

Since $\Delta v$ is the voltage amplitude of a single trap, we can make $%
\delta v=1$ in our calculations or think our calculations scale with this
quantity for the case of one single trap for the sake of simplicity.

In order to analyze the effects of finite size time $\Delta $, to compute $%
var(v)$, we suggest to calculate $var(v)$ for a specific time window that
can change from $\Delta <\tau $, where: $\frac{1}{\tau }=\frac{1}{\tau _{e}}+%
\frac{1}{\tau _{c}}$ up to $\Delta =T$. In our results always $T>>\tau \in
\lbrack \min \left\{ \tau _{e},\tau _{e}\right\} ,\max \left\{ \tau
_{e},\tau _{e}\right\} ]$.

Heuristically, we can obtain $var(v|\Delta )$, the expected variance for an
arbitrary time window of size $\Delta $. One supposes that when $\Delta
<<\tau $, $var(v)\rightarrow 0$, there is no time for occurrences of
captures or emissions. On the other hand, for $\Delta >>\tau $, $%
var(v)\rightarrow var_{\infty }$. Thus we performed a detailed study in
literature, and we tested many functions to perform the transient between
these two regimes. We conclude that transient must behave according to a
sigmoidal function.

The simple sigmoidal behavior, that reproduces these two limits, is the
known Hill-Langmuir \cite{Langmuir,Hill} equation: 
\begin{equation}
f(\Delta )=\frac{\Delta ^{n}}{\Delta ^{n}+\Delta _{c}}\text{,}
\end{equation}%
where $n$ and $\Delta _{c}$ are constants to be fitted. A simple choice is
only to perform the most straightforward form: $n=1$, in order to consider a
function with only available parameter $\Delta _{c}$, that, as we will
observe, has the central role in our modeling.

Thus, aggregating the last information, we obtain the conjecture 
\begin{equation}
var(v|\Delta )=\left\langle v^{2}|\Delta \right\rangle -\left\langle
v|\Delta \right\rangle ^{2}=\frac{\tau _{e}\tau _{c}\left( \tau _{e}+\tau
_{c}\right) ^{-2}}{\left[ 1+\left( \frac{\Delta _{c}}{\Delta }\right) \right]
}\text{,}  \label{Eq:sample_variance}
\end{equation}%
which yields the expected limits: $var(v|0)=0$ e $\lim_{\Delta \rightarrow
\infty }var(v|\Delta )=\frac{\tau _{e}\tau _{c}}{\left( \tau _{e}+\tau
_{c}\right) ^{2}}$, where%
\begin{equation}
\left\langle x|\Delta \right\rangle =xp_{1}(\Delta )+0(1-p_{1}(\Delta
))=xp_{c}(\Delta )\text{.}
\end{equation}%
Thus we hypothesize is that the formulae \ref{Eq:sample_variance} works for
intermediate values of $\Delta $, as we will check. Eq. \ref%
{Eq:sample_variance} also allows, for example, to obtain closed-form
expressions for $p_{0}(\Delta )$ and $p_{1}(\Delta )$, which correspond to
the probabilities of remaining in the states 0 and 1, respectively, as a
function of $\Delta $, just making:

\begin{equation}
p_{1}(\Delta )\left[ 1-p_{1}(\Delta )\right] =\dfrac{\tau _{e}\tau
_{c}\left( \tau _{e}+\tau _{c}\right) ^{-2}}{\left[ 1+\left( \frac{\Delta
_{c}}{\Delta }\right) \right] }\text{.}  \label{Eq:pc}
\end{equation}

Another hypothesis that seems very reasonable, and which we will verify in
this paper, is to consider that $\Delta _{c}$ linearly scales with $\tau $,
i.e., $\Delta _{c}=b\tau $, where $b$ is a constant. Thereby, under these
considerations, solving the equation \ref{Eq:pc}, one has 
\begin{equation}
p_{1}(\Delta )=\frac{1}{2\left( \tau _{e}+\tau _{c}\right) }\left( \tau
_{e}+\tau _{c}\pm \sqrt{\frac{1}{(1+b\frac{\tau }{\Delta })}(\tau
_{e}^{2}+\tau _{c}^{2}-2\tau _{e}\tau _{c}+\frac{b\tau \tau _{e}^{2}+b\tau
\tau _{c}^{2}+2b\tau \tau _{e}\tau _{c}}{\Delta })}\right)  \label{Eq:pc_2}
\end{equation}

The signal says a important point, if $\tau _{e}>\tau _{c}$, the negative
signal leads to $\lim_{\Delta \rightarrow \infty }p_{1}(\Delta )=\frac{\tau
_{c}}{\tau _{c}+\tau _{e}}$, while if $\tau _{c}>\tau _{e}$, the positive
signal leads to $\lim_{\Delta \rightarrow \infty }p_{1}(\Delta )=\frac{\tau
_{c}}{\tau _{c}+\tau _{e}}$, and sure if $\tau _{c}=\tau _{e}$, one has $%
\lim_{\Delta \rightarrow \infty }p_{1}(\Delta )=\frac{1}{2}$. It is also
important to observe that $\lim_{\Delta \rightarrow 0}p_{1}(\Delta )=0$ for
the negative and $\lim_{\Delta \rightarrow 0}p_{1}(\Delta )=1$ for positive,
which makes sense since the state can start with 0 or 1.

Since we understand this semi-analytical treatment, we can imagine
estimating $var(\Delta )$, from a computational/numerical point of view,
which we performed according to sample variance for a specific ($k$-th) time
window by 
\begin{equation}
\widehat{var}_{k}(v|\Delta )=\frac{1}{\Delta -1}\sum_{i=k\Delta
+1}^{(k+1)\Delta }(s_{i}-\overline{s}^{(k)})^{2}
\end{equation}%
where $s_{i}=0,1$ is Bernoulli random variable, and $\overline{s}^{(k)}=%
\frac{1}{\Delta }\sum_{i=k\Delta +1}^{(k+1)\Delta }s_{i}$. Thus, since we
sliced $T$ in $\left\lfloor T/\Delta \right\rfloor $ time windows of size $%
\Delta $, computing the average variance: 
\begin{equation}
\widehat{var}(v|\Delta )=\frac{1}{\left\lfloor T/\Delta \right\rfloor }%
\sum_{k=1}^{\left\lfloor T/\Delta \right\rfloor }var_{k}(v|\Delta )\text{,}
\label{Eq:average_variance}
\end{equation}%
which is a more refined estimate for the theoretical value expressed by Eq. %
\ref{Eq:sample_variance}:

And about the dispersion of $var_{k}(v|\Delta )$? We can numerically
determine it by computing

\begin{equation}
\widehat{var}(var_{k}(v|\Delta ))=\frac{1}{\left\lfloor T/\Delta
\right\rfloor -1}\sum_{k=1}^{\left\lfloor T/\Delta \right\rfloor }\left(
var_{k}(v|\Delta )-\widehat{var}(v|\Delta )\right) ^{2}\text{.}
\label{Eq:var_var_sample}
\end{equation}

However, the estimation theory suggests that an estimator for $\widehat{var}%
(var_{k}(\Delta ))$ has the form: $2\frac{var^{2}(v|\Delta )}{\Delta -1}$
(see for example \cite{Trivedi}). However considering that some adittional
\textquotedblleft ingredients\textquotedblright\ is necessary, we consider
for convenience: 
\begin{equation}
var(var(v|\Delta ))\approx \frac{K(\tau )var^{2}(v|\Delta )}{\Delta }\text{,}
\end{equation}%
where $K(\tau )$ is a \textquotedblleft ad-hoc\textquotedblright\ constant
that depends only on $\tau $, built-in in our analysis. From semi-analytical
formulae for $var(v|\Delta )$ (Eq. \ref{Eq:sample_variance}), we propose:

\begin{equation}
var(var(v|\Delta ))=K(\tau )\frac{\tau _{e}^{2}\tau _{c}^{2}\left( \tau
_{e}+\tau _{c}\right) ^{-4}}{\Delta \left[ 1+\left( \Delta _{c}/\Delta
\right) \right] ^{2}}\equiv \frac{\tau _{e}^{2}\tau _{c}^{2}K(\tau )\Delta }{%
\left( \tau _{e}+\tau _{c}\right) ^{4}\left[ \Delta +\Delta _{c}\right] ^{2}}%
\text{.}  \label{Eq:var_var_one_trap}
\end{equation}%
We can verify that such quantity has a maximum in $\frac{d\ }{d\Delta }%
var(var(v|\Delta ))=0$, resulting in$\frac{1}{\left( \Delta +\Delta
_{c}\right) ^{2}}-\frac{2\Delta }{\left( \Delta +\Delta _{c}\right) ^{3}}=0$%
, and yielding $\Delta _{\max }=\Delta _{c}$, which behaves as $b\tau $ (a
hypothesis to be checked in the section corresponding to our results).

At this point, it is important to highlight some considerations. If this
approach is correct, it has important implications. From experimental
results we can experimentally determine the parameter $\widehat{var}_{\infty
}$ which is numerically equal $\dfrac{\tau _{e}\tau _{c}}{\left( \tau
_{e}+\tau _{c}\right) ^{2}}$, and the parameter $\widehat{\Delta }_{c}$,
that is, according with our modeling, numerically equal to $b\frac{\tau
_{e}\tau _{c}}{\left( \tau _{e}+\tau _{c}\right) }$. Solving these
equations, we can obtain, directly and independently $\tau _{c}$ and $\tau
_{e}$, which are the constants that characterize the noise:

\begin{equation}
\tau _{c}=\frac{\widehat{\Delta }_{c}}{2b\widehat{var}_{\infty }}(1\pm \sqrt{%
1-4\widehat{var}_{\infty }})  \label{Eq:tauc}
\end{equation}%
and

\begin{equation}
\tau _{e}=\frac{\widehat{\Delta }_{c}}{2b\widehat{var}_{\infty }}(1\mp \sqrt{%
1-4\widehat{var}_{\infty }})  \label{Eq:taue}
\end{equation}%
as a function only on experimental parameters:$\ \widehat{\Delta }_{c}$ and $%
\widehat{var}_{\infty }$ respectively given by the inflection point and
steady-state value of noise variance. Observe that the equations \ref%
{Eq:tauc} and \ref{Eq:taue} also clearly show the symmetry between these two
constants. In this case, the only necessary task is to numerically show our
assumptions in the section corresponding to our results. In the following
subsection, we performed an analysis of variance, and variance of variance,
considering a situation of a large number of traps.

\subsection{Analysis with many traps}

In this case, we can assume that $V=\sum_{i=1}^{N_{tr}}v_{i}$ is the
contribution from $N_{tr}$ traps for the voltage, thus

\begin{equation}
\begin{array}{lll}
var(V|\Delta ) & = & var(\sum_{i=1}^{N_{tr}}v_{i}|\Delta ) \\ 
&  &  \\ 
& = & \left\langle \left. \left( \sum_{i=1}^{N_{tr}}v_{i}\right)
^{2}\right\vert \Delta \right\rangle -\left\langle \left. \left(
\sum_{i=1}^{N_{tr}}v_{i}\right) \right\vert \Delta \right\rangle ^{2} \\ 
&  &  \\ 
& = & \sum_{i=1}^{N_{tr}}\left( \left\langle v_{i}^{2}|\Delta \right\rangle
-\left\langle v_{i}|\Delta \right\rangle ^{2}\right) +\sum_{i\neq j}\left(
\left\langle v_{i}v_{j}|\Delta \right\rangle -\left\langle v_{i}|\Delta
\right\rangle \left\langle v_{j}|\Delta \right\rangle \right) \\ 
&  &  \\ 
& = & \sum_{i=1}^{N_{tr}}\left( \left\langle v_{i}^{2}|\Delta \right\rangle
-\left\langle v_{i}|\Delta \right\rangle ^{2}\right) \\ 
&  &  \\ 
& = & \sum_{i=1}^{N_{tr}}var(v_{i}|\Delta )\text{,}%
\end{array}%
\end{equation}%
since the traps are supposedly uncorrelated: $\left\langle v_{i}v_{j}|\Delta
\right\rangle =\left\langle v_{i}|\Delta \right\rangle \left\langle
v_{j}|\Delta \right\rangle $.

Now it is important to consider that $\tau _{c}$ and $\tau _{e}$ are written
as \cite{Kirton}:%
\begin{eqnarray}
\tau _{c} &=&10^{p}(1+e^{q}) \\
&&  \notag \\
\tau _{c} &=&10^{p}(1+e^{-q}),  \notag
\end{eqnarray}%
where $q$ and $p$ are uniformly distributed. Here $q\in \lbrack -Q,Q]$,
while $p\in \lbrack p_{\min },p_{\max }]$, where in this paper we use $Q=2$
and $p_{\min }=1$, and $p_{\max }=7$ which are values experimentally
plausible.

Defining 
\begin{equation}
\left\langle x\right\rangle _{q,p,\Delta v}=\frac{1}{2Q(p_{\max }-p_{\min })}%
\int_{0}^{\infty }\int_{-Q}^{Q}\int_{p_{\min }}^{p_{\max }}w(\delta
v)x(p,q)dpdqd(\delta v),
\end{equation}%
where $w(\delta v)$ is the probability density function of the threshold
voltage of the traps.

We can calculate the average variance considering the contribution of $%
N_{tr} $ traps: 
\begin{equation}
\begin{array}{lll}
\left\langle var(V|\Delta )\right\rangle _{q,p,\delta v} & = & 
\sum_{i=1}^{N_{tr}}\left\langle var(v_{i}|\Delta )\right\rangle _{q,p,\delta
v} \\ 
&  &  \\ 
& = & N_{tr}\left\langle var(v|\Delta )\right\rangle _{q,p,\delta v} \\ 
&  &  \\ 
& = & \frac{N_{tr}\left\langle (\delta v)^{2}\right\rangle }{4Q(p_{\max
}-p_{\min })}\int_{-Q}^{Q}\frac{dq}{(1+\cosh q)}\int_{p_{\min }}^{p_{\max }}%
\frac{dp}{1+\frac{b}{\Delta }10^{p}} \\ 
&  &  \\ 
& = & \frac{N_{tr}\left\langle (\delta v)^{2}\right\rangle }{2Q(p_{\max
}-p_{\min })}\tanh \left( \frac{Q}{2}\right) \left[ (p_{\max }-p_{\min })-%
\frac{1}{\ln 10}\ln \left( \frac{\Delta +b10^{p_{\max }}}{\Delta
+b10^{p_{\min }}}\right) \right]%
\end{array}
\label{Eq:var_many_traps_Ntr}
\end{equation}%
with 
\begin{equation}
\left\langle (\delta v)^{2}\right\rangle =\int_{0}^{\infty }(\delta
v)^{2}w(\delta v)d(\delta v)
\end{equation}

You also can consider that the number of traps in a sample follows a poisson
distribution $p_{\lambda }(N_{tr})=\lambda ^{N_{tr}}\frac{e^{-\lambda }}{%
N_{tr}!}$with rate $\overline{N_{tr}}=\sum_{N_{tr}=0}^{\infty
}N_{tr}p(N_{tr})$. And in this case the variance considering an average over
many samples, is presented only by changing $N_{tr}$ by $\overline{N_{tr}}$: 
\begin{equation}
\overline{\left\langle var(V)\right\rangle _{q,p,\delta v}}=\frac{%
N_{dec}\left\langle (\delta v)^{2}\right\rangle }{2Q}\tanh \left( \frac{Q}{2}%
\right) \left[ (p_{\max }-p_{\min })-\frac{1}{\ln 10}\ln \left( \frac{\Delta
+b10^{p_{\max }}}{\Delta +b10^{p_{\min }}}\right) \right] \text{,}
\label{Eq:var_many_traps_aver_Ntr}
\end{equation}%
where $N_{dec}=\overline{N_{tr}}/(p_{\max }-p_{\min })$.

In this case, MC simulations for a fixed and large $N_{tr}$ substituted in
this equation by $\overline{N_{tr}}$ must corroborate this semi-analytical
result. Similarly from our previous calculations we calculated the variance
of variance from a superposition of $N_{tr}$ traps:

\begin{equation}
\begin{array}{lll}
var(var(V|\Delta )) & = & \left\langle \left(
var(\sum_{i=1}^{N_{tr}}v_{i}|\Delta )\right) ^{2}\right\rangle _{\Delta
}-\left\langle var(\sum_{i=1}^{N_{tr}}v_{i}|\Delta )\right\rangle _{\Delta
}^{2} \\ 
&  &  \\ 
& = & \sum_{i=1}^{N_{tr}}\left\langle var(v_{i}|\Delta )^{2}\right\rangle
_{\Delta }-\left\langle var(v_{i}|\Delta )\right\rangle _{\Delta }^{2}+ \\ 
&  &  \\ 
&  & \sum_{i\neq j}^{N_{tr}}\left\langle var(v_{i}|\Delta )var(v_{j}|\Delta
)\right\rangle _{\Delta }-\left\langle var(v_{i}|\Delta )\right\rangle
\left\langle var(v_{j}|\Delta )\right\rangle _{\Delta } \\ 
&  &  \\ 
& = & \sum_{i=1}^{N_{tr}}\left\langle var(v_{i}|\Delta )^{2}\right\rangle
_{\Delta }-\left\langle var(v_{i}|\Delta )\right\rangle _{\Delta }^{2} \\ 
&  &  \\ 
& = & \sum_{i=1}^{N_{tr}}var(var(v_{i}|\Delta ))%
\end{array}%
\end{equation}%
since $\left\langle var(v_{i}|\Delta )var(v_{j}|\Delta )\right\rangle
_{\Delta }=\left\langle var(v_{i}|\Delta )\right\rangle \left\langle
var(v_{j}|\Delta )\right\rangle _{\Delta }$. The problem here stays in our
ignorance about $K(\tau )$ in Eq. \ref{Eq:var_var_one_trap}. The simple
choice is simply to suppose a linear dependence: $K(\tau )=\gamma \tau
=\gamma 10^{p}$, and following exactly as

\begin{equation}
\begin{array}{ccc}
\overline{\left\langle var(var(V|\Delta ))\right\rangle _{q,p,\Delta v}} & =
& \frac{N_{dec}\gamma \Delta }{4Q}\left\langle (\delta v)^{4}\right\rangle
\int_{-Q}^{Q}\frac{dq}{(1+\cosh q)^{2}}\int_{p_{\min }}^{p_{\max }}\frac{%
10^{p}}{\left[ \Delta +b10^{p}\right] ^{2}}dp \\ 
&  &  \\ 
& = & \frac{N_{dec}\gamma \Delta }{6Q\ln 10}\left\langle (\delta
v)^{4}\right\rangle \frac{\sinh Q(\cosh Q+2)}{(\cosh Q+1)^{2}}\cdot \frac{%
10^{p_{\max }}-10^{p_{\min }}}{(b10^{p_{\max }}+\Delta )(b10^{p_{\min
}}+\Delta )}%
\end{array}
\label{Eq:many_traps_var_var}
\end{equation}%
where $\left\langle (\delta v)^{4}\right\rangle =\int_{0}^{\infty }(\delta
v)^{4}w(\delta v)d(\delta v)$.

We will present our results divided into two parts: I -- \textit{Single trap}
and II -- \textit{Many traps}, presented in the following.

\section{ Results Part I: Single trap}

\begin{figure}[tbp]
\begin{center}
\includegraphics[width=0.5\columnwidth]{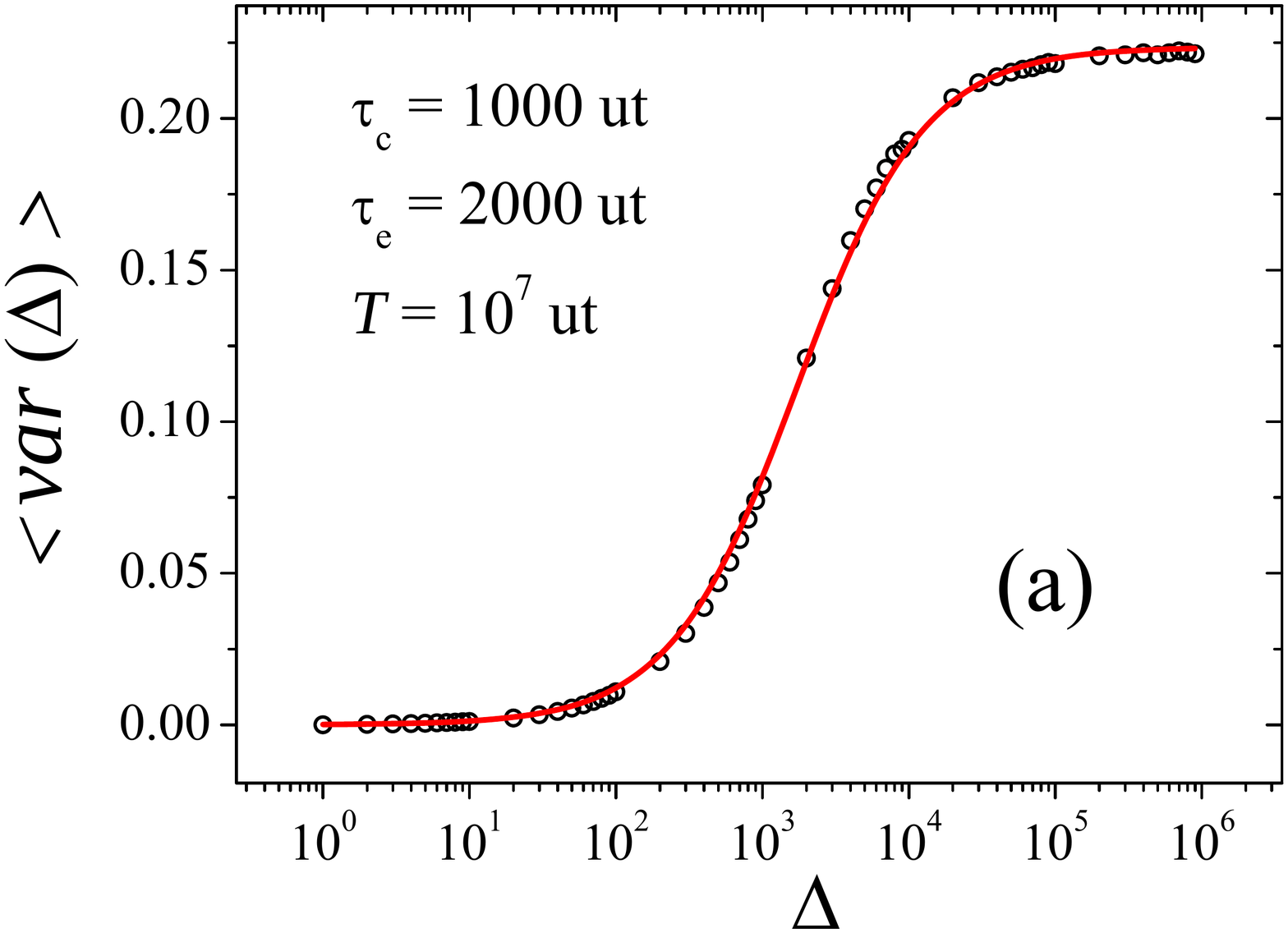}\includegraphics[width=0.5%
\columnwidth]{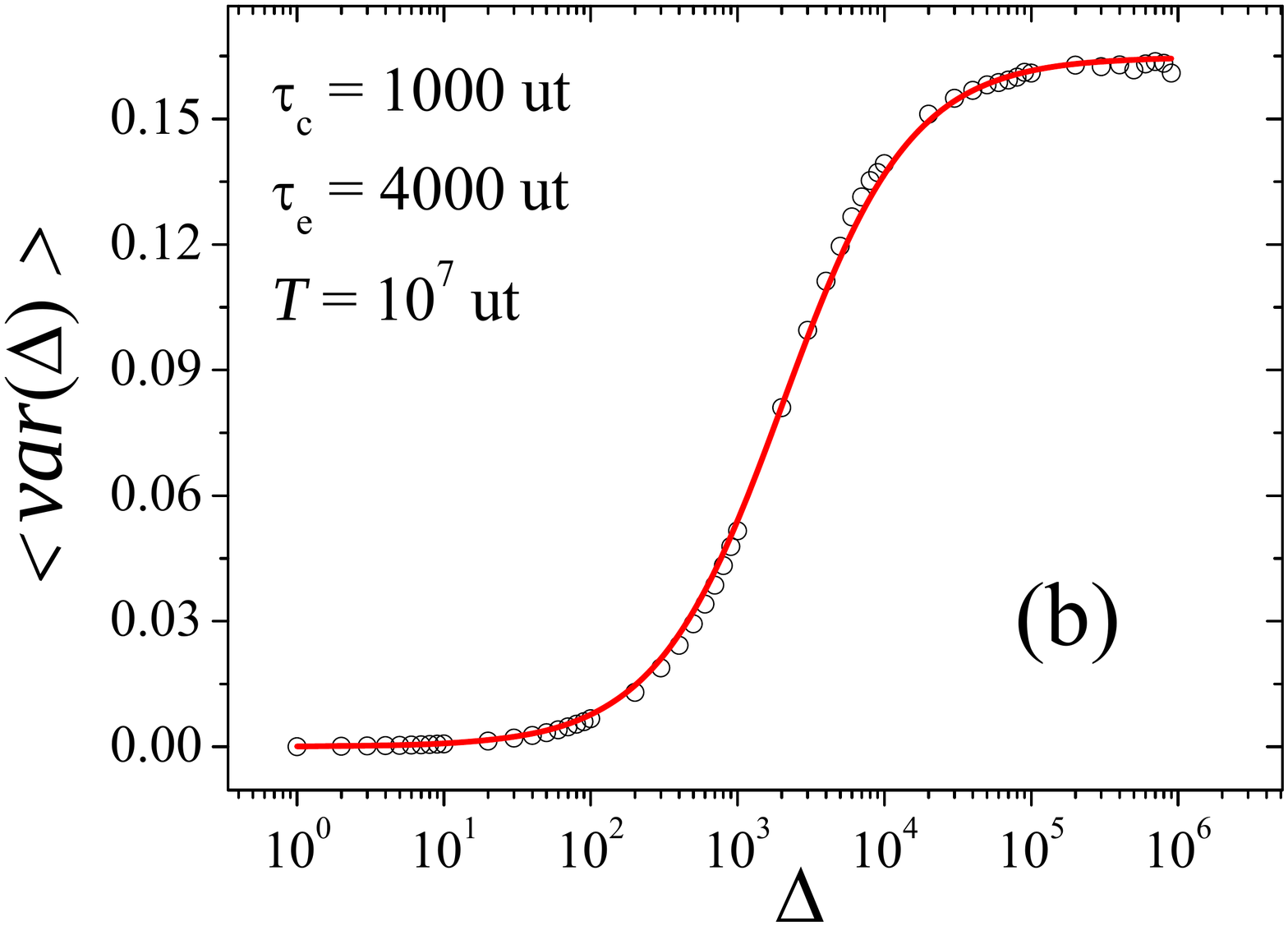}
\end{center}
\caption{(a) The verage variance as a function of the time window. Here, we
used $\protect\tau _{c}=1000$ and $\protect\tau _{e}=2000$. (b) The average
variance as a function of time window when $\protect\tau _{c}=1000$ and $%
\protect\tau _{e}=4000$. Points correspond to MC simulations while the red
line corresponds to our semi-analytical result according to Eq. \protect\ref%
{Eq:sample_variance}}
\label{Fig:average_variance}
\end{figure}

Let us start with simple experiments that corroborate our semi-analytical
close-form expressions from Eqs. \ref{Eq:sample_variance} and \ref%
{Eq:var_var_one_trap}. Keeping $\tau _{c}=1000$ u.t. (unit of time), we
performed MC simulations changing $\tau _{e}$. In these simulations we used
a total of $T=10^{7}$ u.t. In our simulations as previously reported 1 u.t =
1 MC step. In Fig. \ref{Fig:average_variance} we show the sample variance
for the particular cases $\tau _{e}=2000$ u.t and $\tau _{e}=4000$ u.t. We
show the MC simulations (points) according to Eq. \ref{Eq:average_variance}
compared with our semi-analytical result (Eq. \ref{Eq:sample_variance}) red
curve.

We observe a excellent agreement between MC and our semi-analytical result.
It is important to observe that we numerically obtain $\tau _{e}\tau
_{c}\left( \tau _{e}+\tau _{c}\right) ^{-2}\approx 0.223$ for $\tau
_{c}=1000 $ u.t and $\tau _{c}=2000$ u.t in comparison with exact result: $%
1000\times 2000\times \left( 3000\right) ^{-2}=\allowbreak 0.2\overline{2}$
and $\tau _{e}\tau _{c}\left( \tau _{e}+\tau _{c}\right) ^{-2}\approx 0.165$
for $\tau _{c}=1000$ u.t and $\tau _{c}=4000$ u.t, which corroborates the
exact value: $0.16$. The same fits also lead to $\Delta _{c}=1735$ u.t and $%
\Delta _{c}=2050$ u.t respectively.

\begin{figure}[tbp]
\begin{center}
\includegraphics[width=0.5\columnwidth]{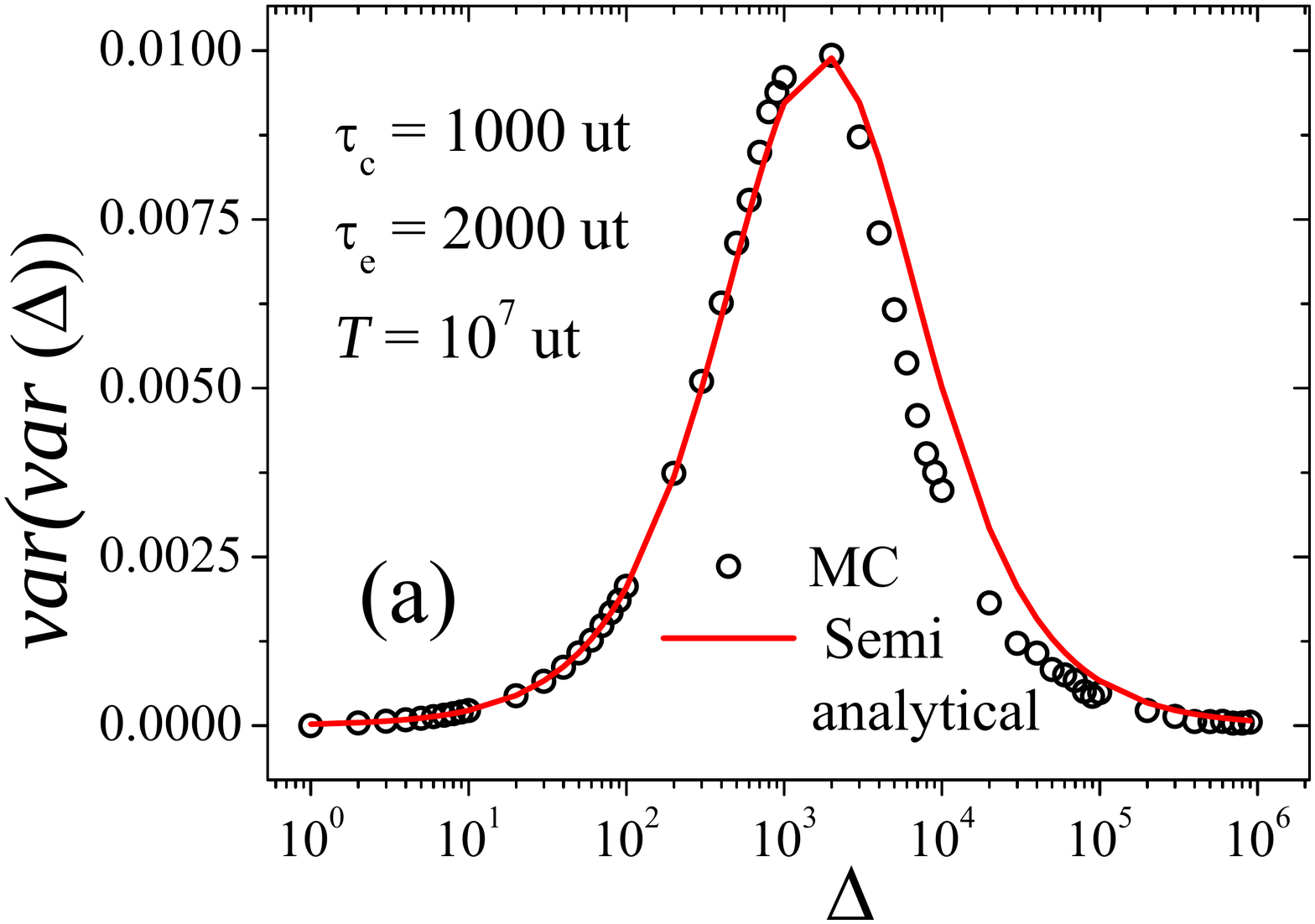}\includegraphics[width=0.5%
\columnwidth]{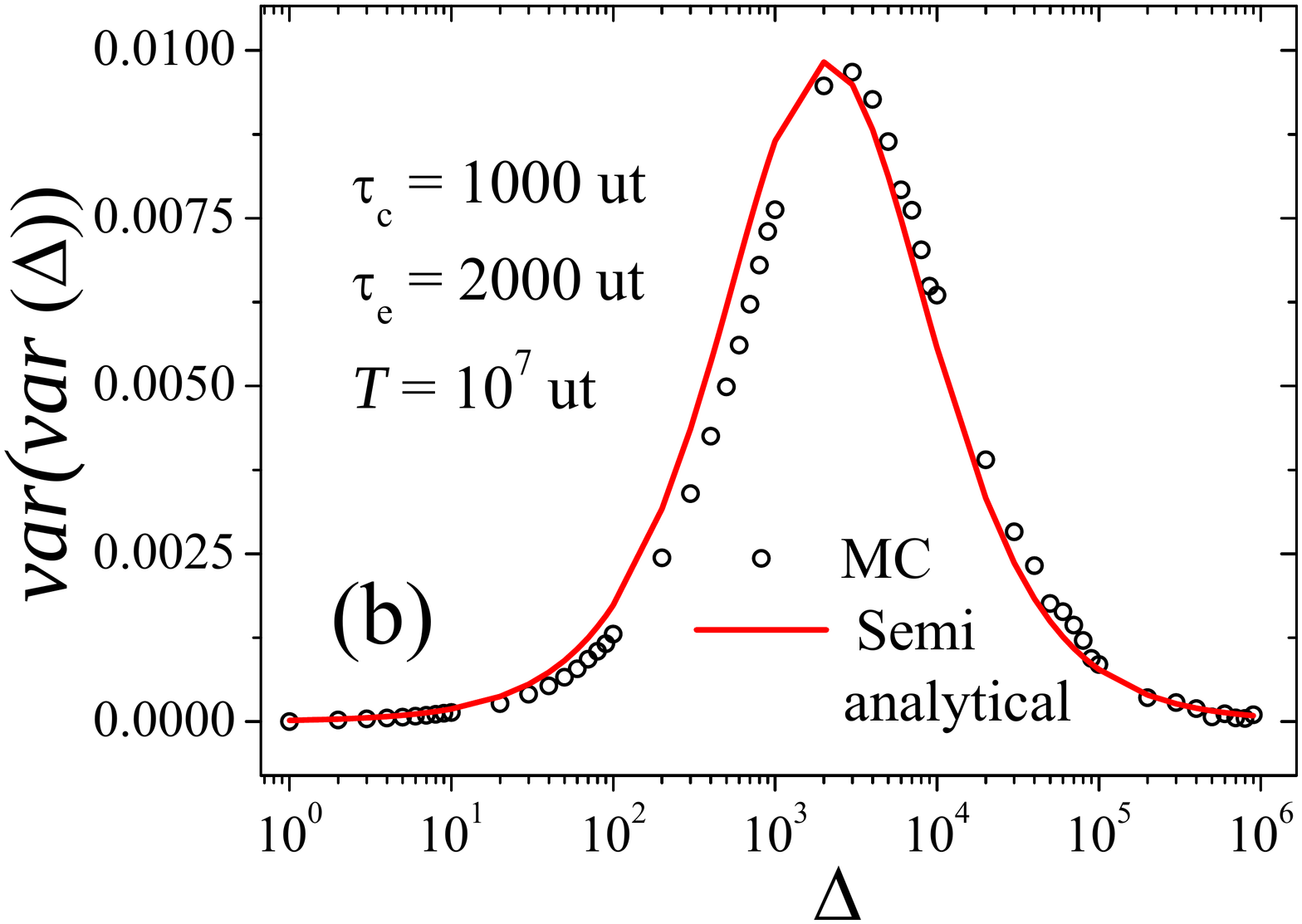}
\end{center}
\caption{(a) The variance of sample variance as a function of time window: $%
\protect\tau _{c}=1000$ and $\protect\tau _{e}=2000$. (b) The variance of
sample variance as a function of time window: $\protect\tau _{c}=1000$ and $%
\protect\tau _{e}=4000$. Points correspond to MC simulations while red line
correspond to fit with Eq. \protect\ref{Eq:var_var_one_trap}}
\label{Fig:variance_of_variance}
\end{figure}

In order to check such estimates, we perform MC simulations by calculating $%
var(var_{k}(\Delta ))$ numerically by Eq. \ref{Eq:var_var_sample}. Thus we
fit these results with function 
\begin{equation}
f(\Delta )=k_{\max }\frac{\Delta }{\left( \Delta +\Delta _{c}\right) ^{2}}
\label{Fig:Fit_for_variance_of_variance}
\end{equation}%
according to predicted by Eq. \ref{Eq:var_var_one_trap}. Here $k_{\max }$ is
a constant only to adjust the high of the peak. Thus, using the values of $%
\Delta _{c}$ we can observe the good agreement between MC simulations
(points) and our semi-analytical prediction (red continuous curve) in Fig. %
\ref{Fig:variance_of_variance}.

\begin{figure}[tbp]
\begin{center}
\includegraphics[width=0.8\columnwidth]{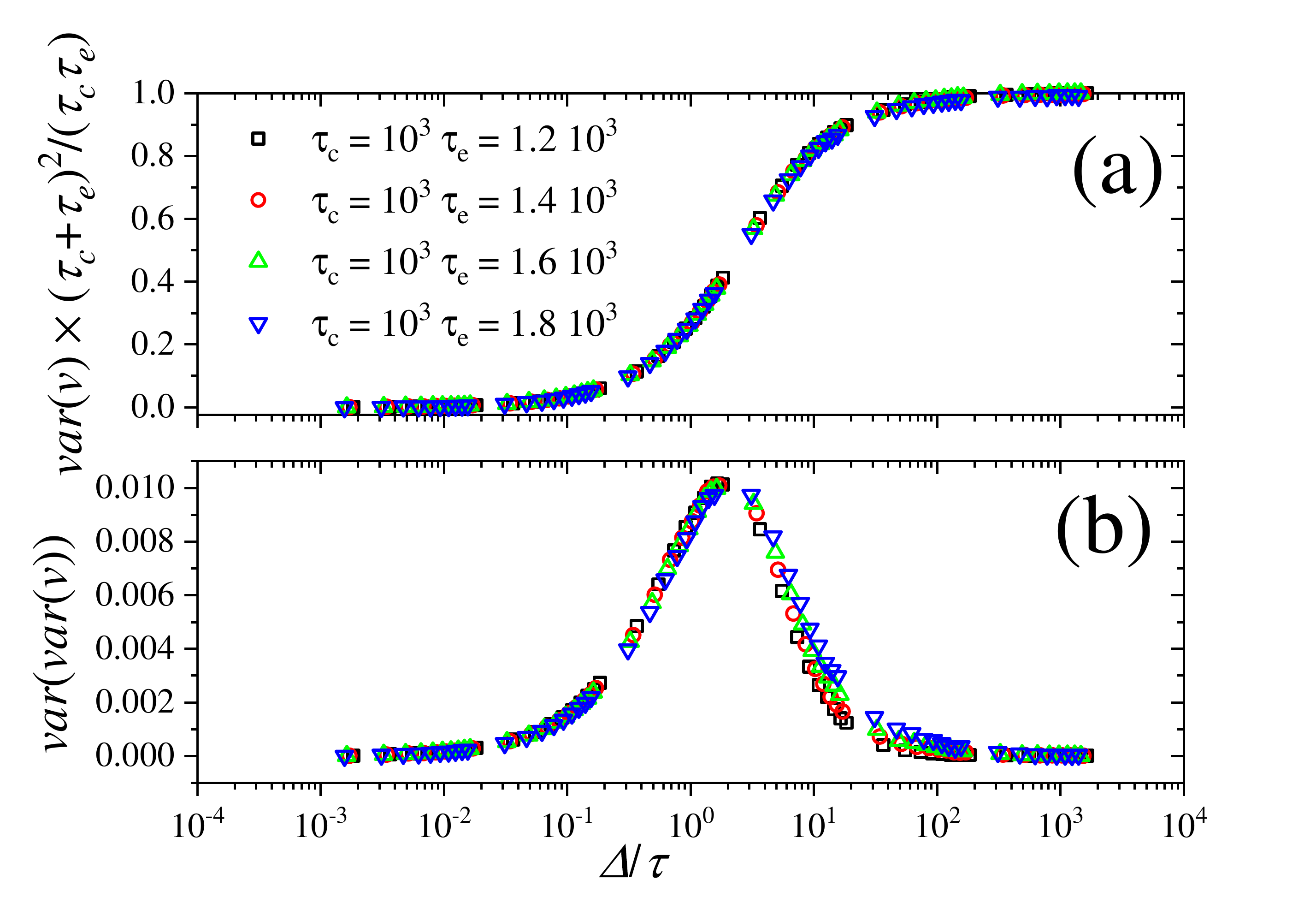}
\end{center}
\caption{Collapse for different values of $\protect\tau _{e}$ and $\protect%
\tau _{c}$. (a) Average of the sample variance (b) Variance of the variance. 
}
\label{Fig: collapse}
\end{figure}

We observe the collapse of curves for different $\tau _{c}$ and $\tau _{e}$
if we divide the time window by $\tau $ and if we multiply the noise
variance by $(\tau _{c}+\tau _{e})^{2}/\tau _{c}\tau _{e}$. The plots for
the average variance and sample variance of variance are shown respectively
in Fig. \ref{Fig: collapse}.

\begin{figure}[tbp]
\begin{center}
\includegraphics[width=0.7\columnwidth]{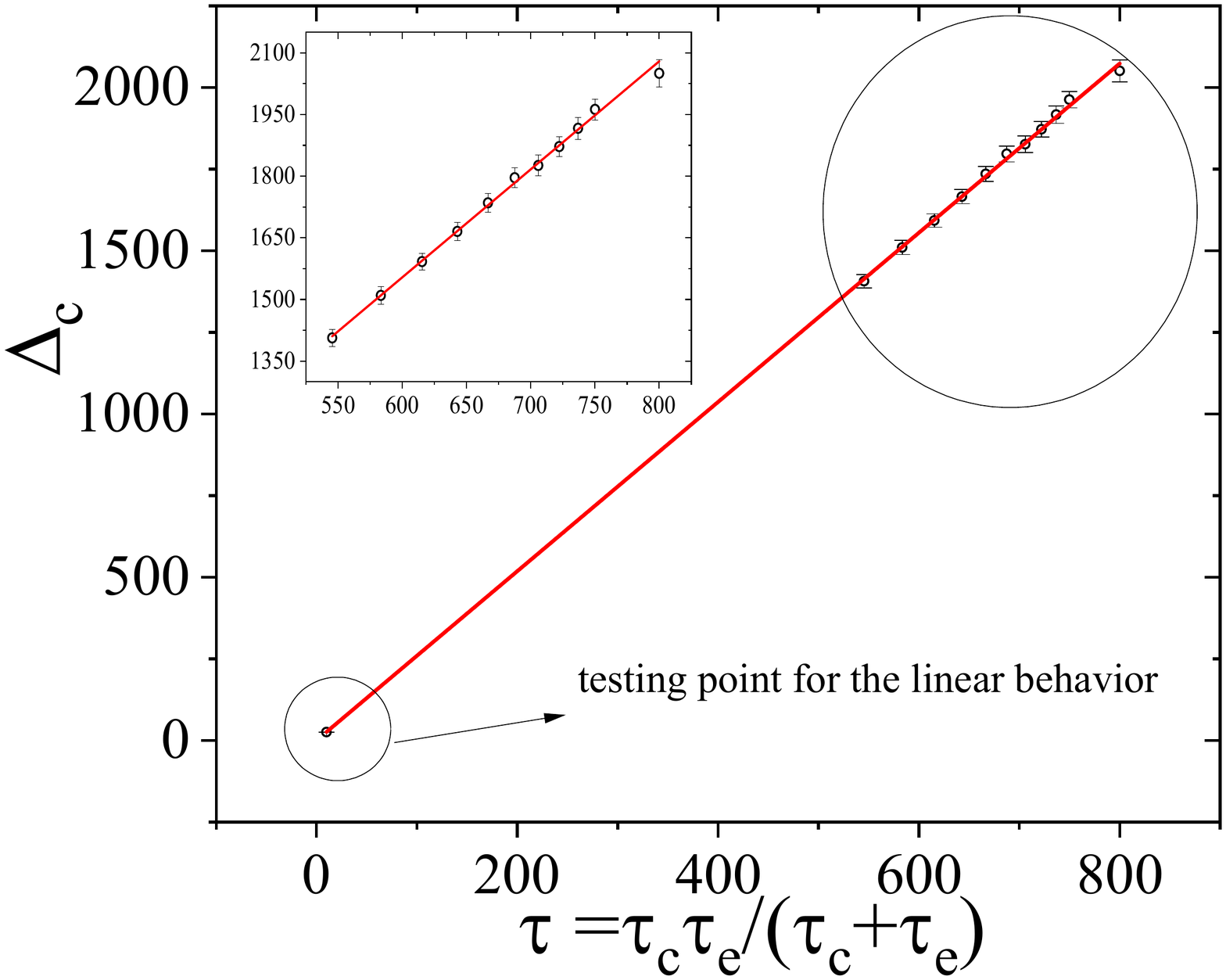}
\end{center}
\caption{Linear behavior of $\Delta _{c}$ as a function of $\protect\tau $.
We use a testing point to show that linear behavior is robust for an
extensive range in $\protect\tau $. The inset plot is only a highlight for
the selected region. }
\label{Fig:deltac_versus_deltae-deltac}
\end{figure}

To complete the validation of the process, which allows for example that we
can calculate $\tau _{c}$ and $\tau _{e}$ from experimental results through
Eqs. \ref{Eq:tauc} and \ref{Eq:taue}, we need to establish how $\Delta _{c}$
depends on $\tau _{c}$ and $\tau _{e}$. More precisely, we need to show our
assumption in this paper that is $\Delta _{c}=b\tau $. Thus considering $%
\tau _{c}=1000$ we change $\tau _{e}$, for that pair $\tau _{c}$ and $\tau
_{e}$ we estimate $\Delta _{c}$. Thus we plot $\Delta _{c}$ as function of $%
\tau =\frac{\tau _{e}\tau _{c}}{\left( \tau _{e}+\tau _{c}\right) }$. Fig. %
\ref{Fig:deltac_versus_deltae-deltac} shows a robust linear behavior,
corroborating the hypothesis $\Delta _{c}=b\ \tau $. The inset plot is only
a highlight for the selected region.

The first isolated point (corresponding to $\tau _{e}=10$) in order to show
that the linear behavior is robust indeed. The linear fit leads to $%
b=2.62\pm 0.04$ and corroborates the hypothesis of our modeling for one trap.

\section{ Results Part II: Many traps}

We performed MC simulations considering the contribution of many traps for
the noise. Since that distribution $w(\delta v)$ is not known, we make $%
\left\langle (\delta v)^{2}\right\rangle $ and $\left\langle (\delta
v)^{4}\right\rangle $ identically equal to 1, for the sake of simplicity, or
we should imagine that amounts divided by these moments, which is not a
technical problem here.

\begin{figure}[tbp]
\begin{center}
\includegraphics[width=0.5\columnwidth]{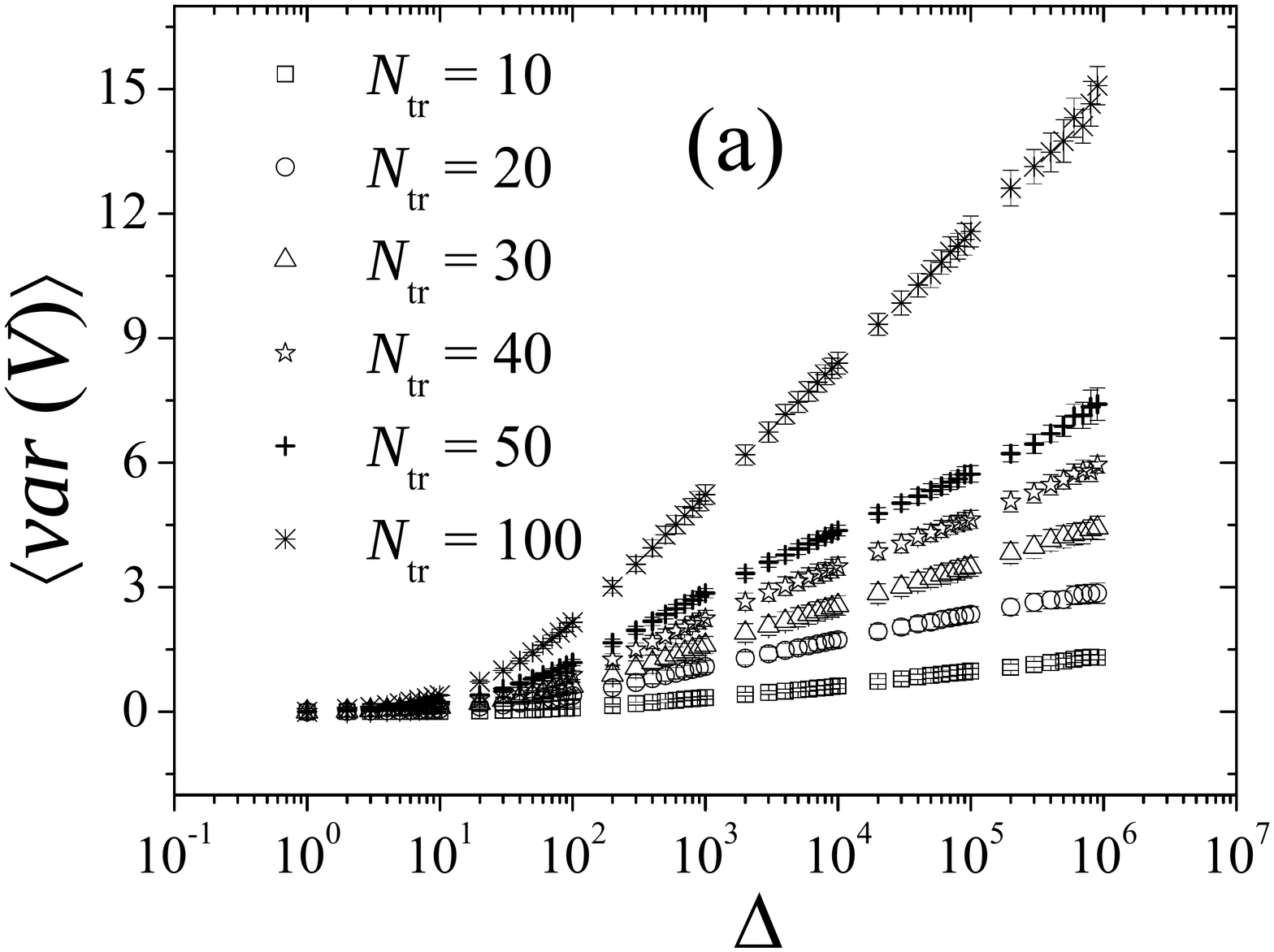}%
\includegraphics[width=0.5\columnwidth]{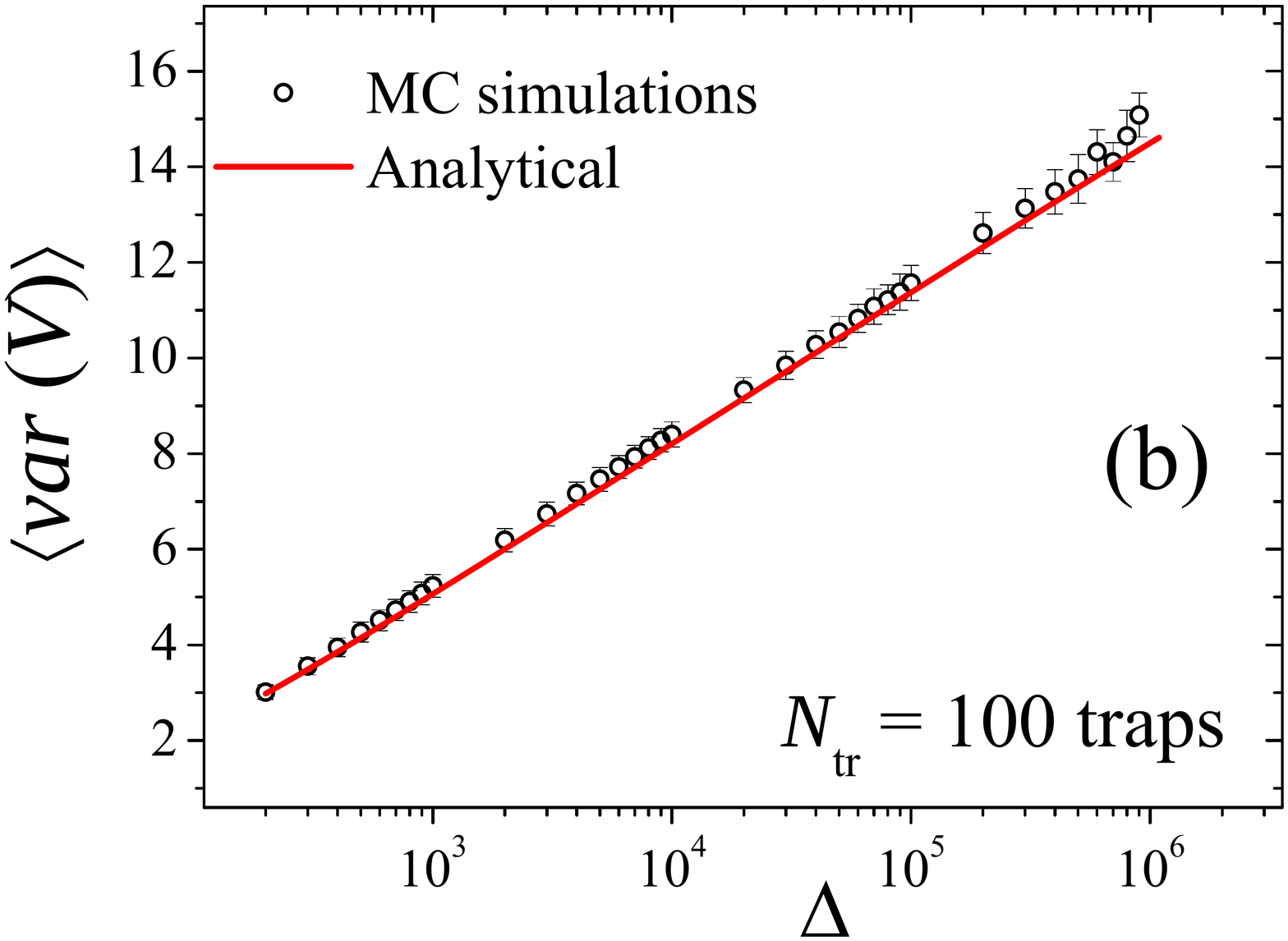}
\end{center}
\caption{(a) The average variance for many traps via MC simulations.
Analysis of the different number of traps (b) Selection of de case of $%
N_{tr}=100$ traps by extracting the first occurrences. We observe an
excellent agreement between MC and our semi-analytical approach. }
\label{Fig:variance_many_traps}
\end{figure}

We considered simulations with $N_{tr}=10$, $20$, $30$, $40$, $50$, and $100$%
. In Figs. \ref{Fig:variance_many_traps} (a) and (b). In Fig. \ref%
{Fig:variance_many_traps} (a) we show our MC simulations for different
number of traps, while in Fig. \ref{Fig:variance_many_traps} (b) a
comparison between MC simulations and our semi-analytical approach given by
Eq. \ref{Eq:var_many_traps_Ntr} is presented for the larger number of traps
studied: $N_{tr}=100$.

We observe an excellent agreement between the MC simulations and our
semi-analytical approach. We can obtain a simple closed-form if we consider
a seemingly crude approximation. Let us imagine a situation of a
intermediate time window, where $\Delta >>10^{p_{\min }}$ but $\Delta
<<10^{p_{\max }}$. In this case 
\begin{equation*}
\ln \left( \frac{\Delta +b10^{p_{\max }}}{\Delta +b10^{p_{\min }}}\right)
\approx \ln b+p_{\max }\ln 10-\ln \Delta
\end{equation*}

And from Eq. \ref{Eq:var_many_traps_Ntr} one has that 
\begin{equation}
\left\langle var(V|\Delta )\right\rangle _{q,p,\delta v}\approx A+B\ln \Delta
\label{Eq:linear_in_log_scale}
\end{equation}%
where $A=\zeta N_{dec}$ and$\ B=\xi N_{dec}$, with 
\begin{equation}
\zeta =-\frac{1}{2Q}\tanh \left( \frac{Q}{2}\right) \left[ p_{\min }+\log b%
\right]
\end{equation}%
and 
\begin{equation}
\xi =\frac{1}{2Q\ln 10}\tanh \left( \frac{Q}{2}\right)  \label{Eq:beta}
\end{equation}

Taking advantage of the results used for the Fig. \ref%
{Fig:variance_many_traps} (a) we perform linear fits of $\left\langle
var(V|\Delta )\right\rangle _{q,p,\delta v}$ as function of $\ln \Delta $
for the different values of $N_{tr}$.

\begin{figure}[tbp]
\begin{center}
\includegraphics[width=1.0\columnwidth]{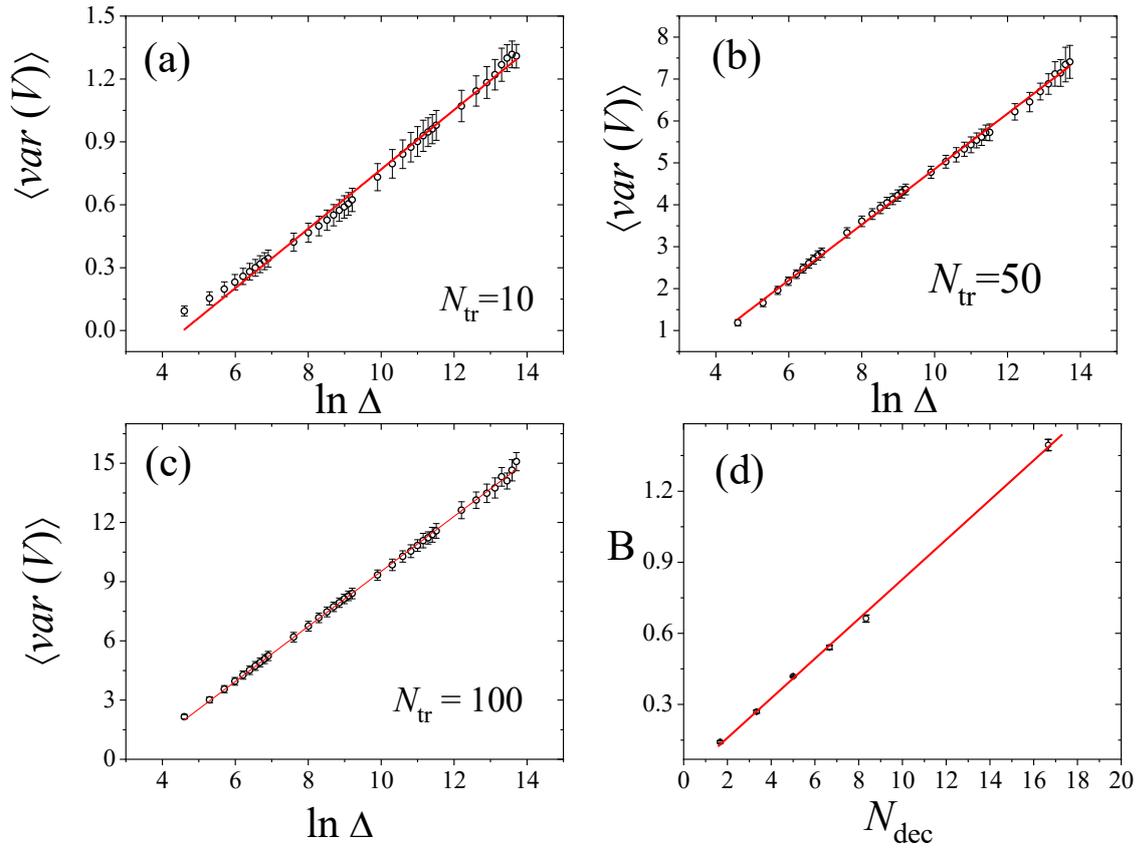}
\end{center}
\caption{(a) The average as a function of $\ln \Delta $ for $N_{tr}=10$. The
continuous curve corresponds to a linear fit. Plots (b) and (c) correspond
to the same one performed in the plot (a) but for $N_{tr}=50$ and $100$
respectively. Finally, plot (d) corresponds to $B$ as a function of $N_{dec}$%
. We also observe an expected linear behavior where the slope corresponds to 
$\protect\xi $. }
\label{Fig:coeficcients}
\end{figure}

Let us concentrate in the slope, i.e., the coefficient $B$ due to its
greater importance. Thus we obtain $B$ for the different values of $N_{tr}$
as shown in Fig. \ref{Fig:coeficcients} (a), (b), and (c) shows the linear
behavior expected by Eq. \ref{Eq:linear_in_log_scale} for respectively $%
N_{tr}=10$, $50$, and $100$. The red continuous curves correspond to linear
fits. The same linear behavior is observed for $N_{tr}=20$, $30$, and $40$
(omitted by saving space).

Fig. \ref{Fig:coeficcients} (d) shows exactly the slope $B$ obtained for
these plots as function of $N_{dec}$. A linear fit also is performed and the
slope must correspond to $\xi $ according to Eq. \ref{Eq:beta}. The slope
leads to: $\xi _{MC}=0.084\,6\pm 0.0012$. Using the values $Q=2$, $p_{\min
}=1$, $p_{\max }=7$, one obtains:

\begin{equation}
\xi _{theor}=\frac{\tanh (1)}{4}\frac{1}{\ln 10}=0.0827,
\end{equation}%
$\,$which agrees with $\xi _{MC}$ considering two uncertainty bars. In
summary, the Eq. \ref{Eq:linear_in_log_scale} is indeed correct by showing
that variance of many traps behaves linearly on the logarithm of time window
in RTS with slope expressed by Eq. \ref{Eq:beta}.

Finally, let us consider the case of the variance of variance considering
the contribution of many traps. We start with a \textquotedblleft
pedagogical\textquotedblright\ situation where we look at the situation of
two and three traps in the sample $(N_{tr}=2$ and $3$).

\begin{figure}[tbp]
\begin{center}
\includegraphics[width=0.8\columnwidth]{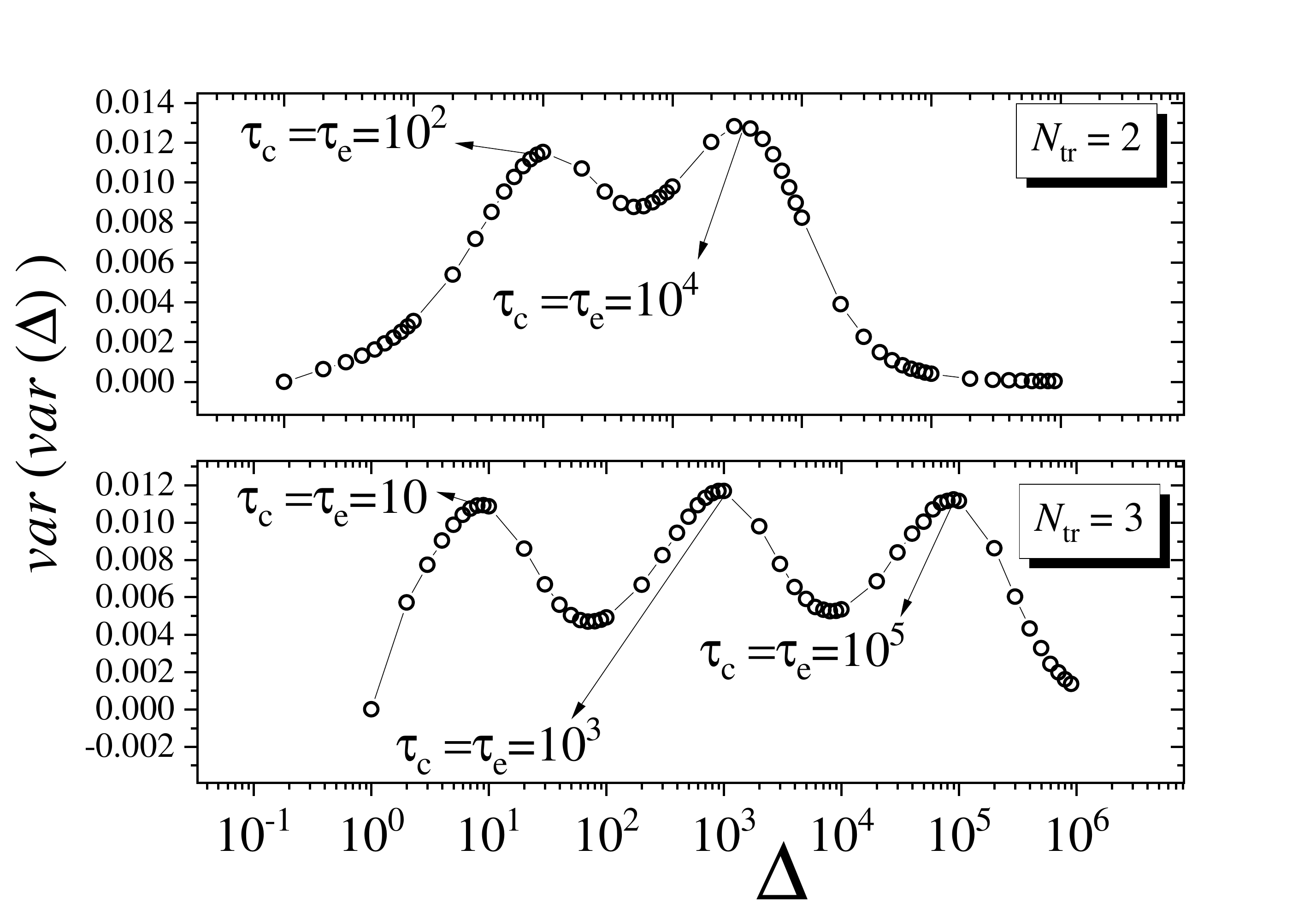}
\end{center}
\caption{The variance of sample variance as a function of the time window
considering (a) $N_{tr}=2$ traps, and (b) $N_{tr}=3$ traps. We averaged over 
$N_{run}=20$ different runs. }
\label{Fig:2_and_3_traps}
\end{figure}
We can observe two peaks and three peaks, respectively, which is intuitive
since for one trap, we observed a peak in $\Delta _{c}=b\tau $, and traps
with different constants lead to peaks in the respective values. And about a
large number of traps? Is the Eq. \ref{Eq:many_traps_var_var} valid? Is it
compatible with MC simulations?

\begin{figure}[tbp]
\begin{center}
\includegraphics[width=0.50\columnwidth]{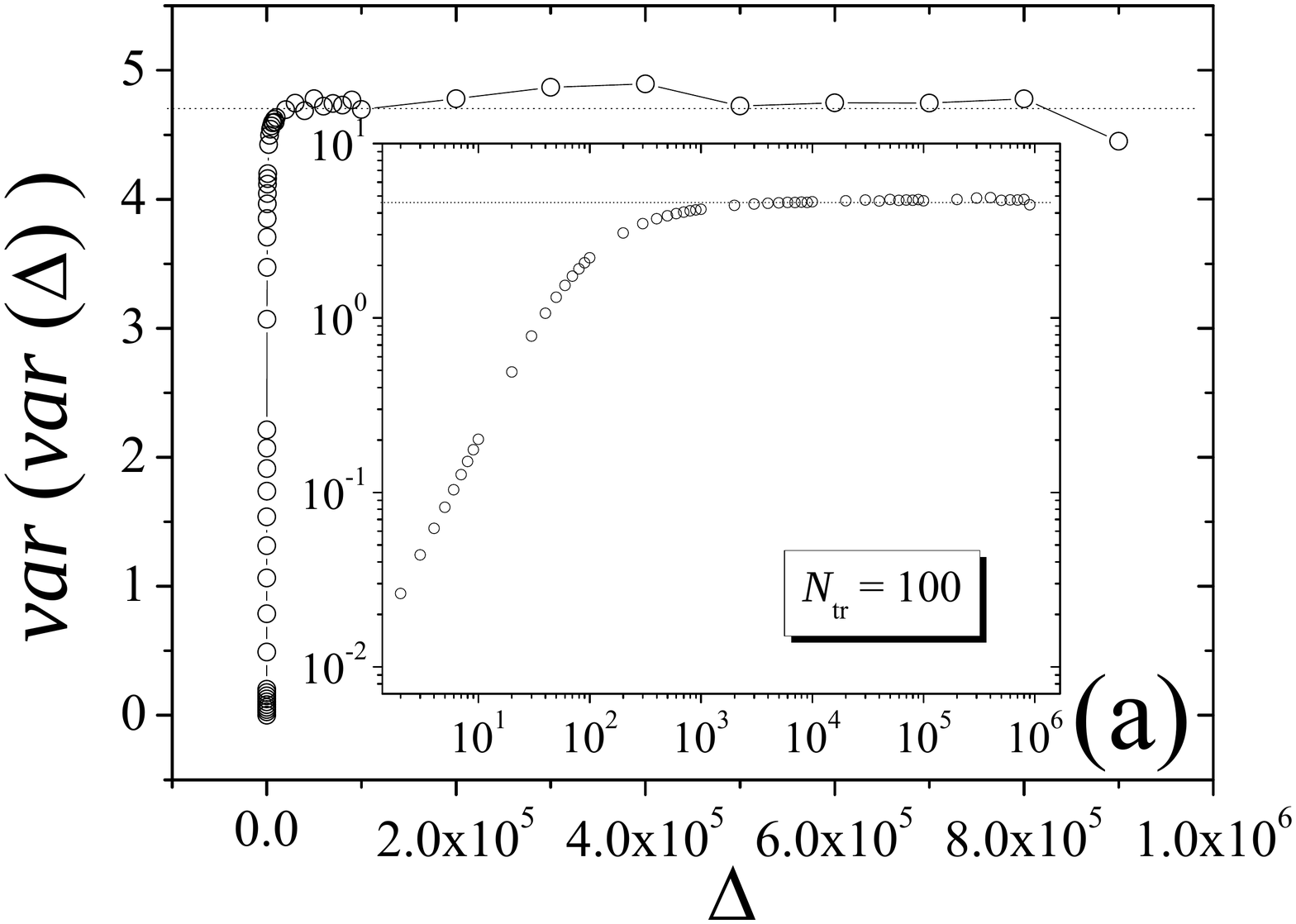}%
\includegraphics[width=0.50\columnwidth]{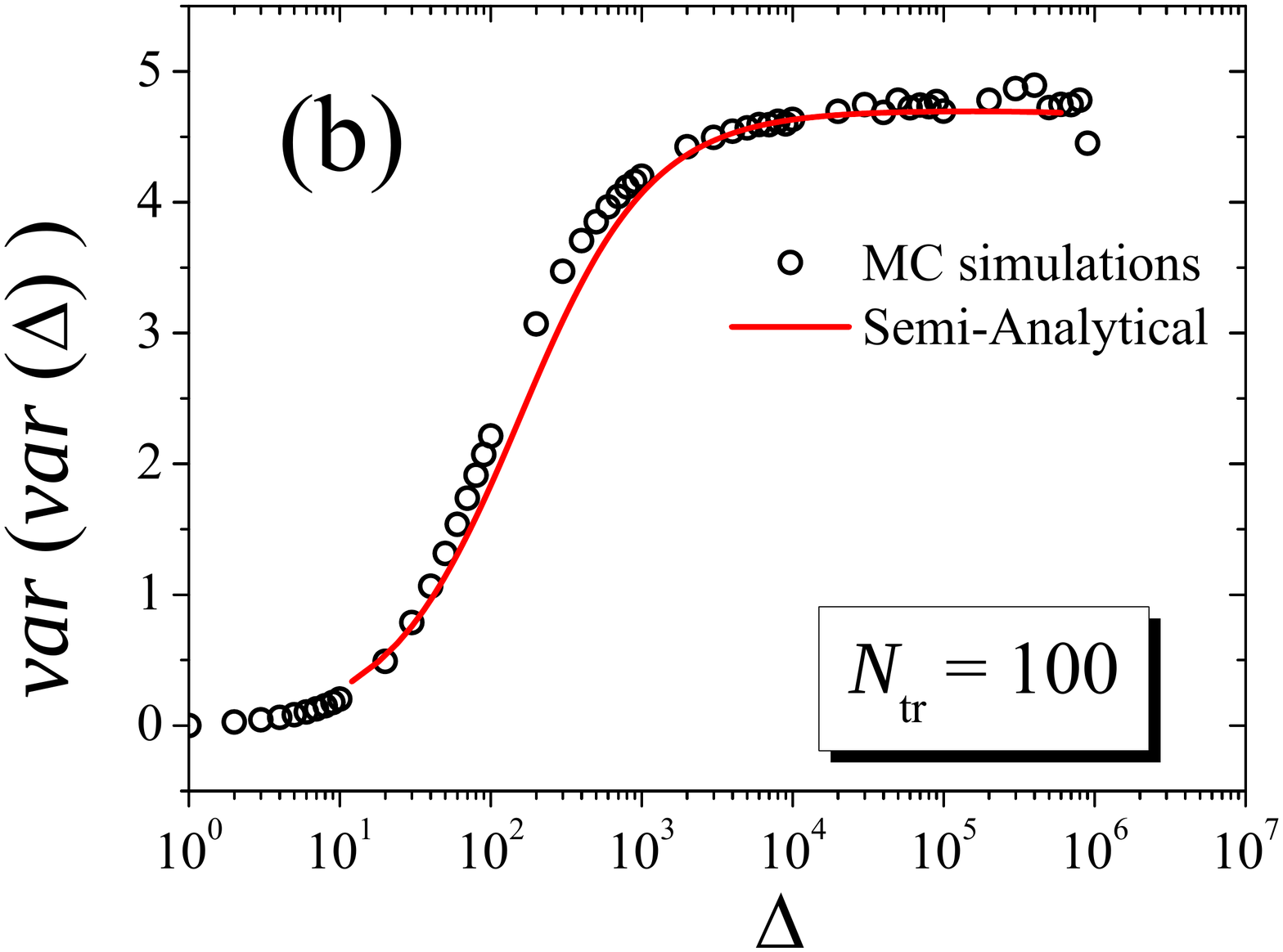}
\end{center}
\caption{The variance of sample variance as a function of the time window
considering many traps: $N_{tr}=100$. Plot (a) describes the situation on a
linear scale. The inset plot corresponds to the same plot on a log-log
scale. Plot (b) considers the same plot with a semi-log scale (log in the
time window axis). We compared MC simulations (points) with the result from
Eq. \protect\ref{Eq:many_traps_var_var}.}
\label{Fig:variance_variance_many_many_traps}
\end{figure}

The variance of variance as a function of the time window considering many
traps is also analyzed. We fixed $N_{tr}=100$. Fig \ref%
{Fig:variance_variance_many_many_traps} (a) describes the plot in linear
scale, which shows that variance of the variance via MC simulations,
considering a contribution of many traps, assumes a constant value in a
region of time windows far from $T$. The inset plot corresponds to the same
plot on a log-log scale. It highlights the constant behavior. Fig \ref%
{Fig:variance_variance_many_many_traps} (b) considers the same plot now in
the scale used in all other cases of this paper: semi-log plot (long in the
time window axis). We compared MC simulations (points) with the result from
Eq. \ref{Eq:many_traps_var_var} showing that both corroborate the constant
behavior for $\Delta <<T$. In this case, once adjusted the constant $\gamma $%
, we considered for the plot that four units of time of our semi-analytical
result (Eq. \ref{Eq:many_traps_var_var}) is equivalent to 1 MC step. The
superposition of the many peaks seems to lead to this constant behavior.
Sure, our plots considered $\Delta _{\max }=10^{5}$ while $T=10^{8}$.

\section{Conclusions}

Our results show universal behavior for the variance and variance of the
sample variance of the RTS noise as a function of the time window for the
analysis of one trap, i.e., supposing that RTS noise is due to a dominant
trap. Our semi-analytical approach agrees with MC simulations and suggests a
simple form to compute the time-constants of the RTS noise. Our analysis for
many traps shows that variance of the RTS noise follows a linear behavior
with the logarithm of the time window, which slope is theoretically
estimated by $\xi _{theor}=$ $\allowbreak 0.0827$. Finally, our results
suggest that variance of variance does not depend on the observation window $%
\Delta $. We are sure that these semi-analytical results, described by
closed-form equations, and corroborated by MC simulations, can be promptly
applied by circuit designers, scientists, and students to project devices or
analyze them.

\section*{Acknowledgements}

R. da Silva would like to thank CNPq for the partial financial support under
the grant: 311236/2018-9

\end{document}